\documentclass[conference]{IEEEtran}
\IEEEoverridecommandlockouts
\usepackage{cite}
\usepackage[acronym]{glossaries}
\usepackage{subcaption}
\usepackage{amsmath,amssymb,amsfonts}
\usepackage{algorithmic}
\usepackage{graphicx}
\usepackage{textcomp}
\usepackage{xcolor}
\usepackage{setspace}

\usepackage{threeparttable}
\usepackage{tabulary}
\usepackage{booktabs}
\usepackage{adjustbox}
\usepackage{url}
\usepackage{xspace}
\usepackage{circledsteps}
\usepackage{multirow}
\usepackage{microtype}
\usepackage{tablefootnote}
\usepackage{enumitem}

\usepackage{siunitx}
\DeclareSIUnit{\nothing}{\relax}
\DeclareSIUnit{\degree}{°}
\DeclareSIUnit{\deg}{deg}
\DeclareSIUnit\pixel{px}
\DeclareSIUnit{\fps}{frame/s}
\DeclareSIUnit{\mac}{MAC}

\definecolor{PULPOrange}{HTML}{F29545}
\definecolor{Turqouise}{HTML}{2CA089}
\definecolor{NetworkMagenta}{HTML}{cb0bc0}

\newcommand\numberincircle[1]{\Circled[inner color=white, outer color=white, fill color=PULPOrange]{#1}}
\newcommand\numberincircletiming[1]{\Circled[inner color=white, outer color=white, fill color=Turqouise]{#1}}

\makeatletter
\newcommand\footnoteref[1]{\protected@xdef\@thefnmark{\ref{#1}}\@footnotemark}
\makeatother

\def\BibTeX{{\rm B\kern-.05em{\sc i\kern-.025em b}\kern-.08em
    T\kern-.1667em\lower.7ex\hbox{E}\kern-.125emX}}
\begin{document}

\setstretch{0.89} 

\bstctlcite{IEEEexample:BSTcontrol}

\newacronym[plural=TNNs, firstplural={Ternary Neural Networks (TNNs)}]{tnn}{TNN}{Ternary Neural Network}
\newacronym{cutie}{CUTIE}{Completely Unrolled Ternary Inference Engine}
\newacronym{soa}{SoA}{State-of-the-Art}
\newacronym{soc}{SoC}{System-on-Chip}
\newacronym{sne}{SNE}{Spiking Neural Engine}
\newacronym{ml}{ML}{Machine Learning}
\newacronym{dnn}{DNN}{Deep Neural Network}
\newacronym{dvs}{DVS}{Dynamic Vision Sensor}
\newacronym{dvss}{DVSs}{Dynamic Vision Sensors}
\newacronym{iot}{IoT}{Internet of Things}
\newacronym[plural=SNNs, firstplural={Spiking Neural Networks (SNNs)}]{snn}{SNN}{Spiking Neural Network}
\newacronym[plural=UAVs, firstplural={unmanned aerial vehicles (UAVs)}]{uav}{UAV}{unmanned aerial vehicle}

\title{Circuits and Systems for Embodied AI: Exploring uJ Multi-Modal Perception for Nano-UAVs on the Kraken Shield}
\author{
\IEEEauthorblockN{Viviane Potocnik\IEEEauthorrefmark{2}, 
Alfio Di Mauro\IEEEauthorrefmark{2}, 
Lorenzo Lamberti\IEEEauthorrefmark{3}, 
Victor Kartsch\IEEEauthorrefmark{2}, 
Moritz Scherer\IEEEauthorrefmark{2}, 
Francesco Conti\IEEEauthorrefmark{3}, \\ 
Luca Benini\IEEEauthorrefmark{2}\IEEEauthorrefmark{3}}
\IEEEauthorblockA{\IEEEauthorrefmark{2} ETH Z\"{u}rich, Switzerland, \IEEEauthorrefmark{3} University of Bologna, Italy }}%
\maketitle

\begin{abstract}
    Embodied artificial intelligence (AI) requires pushing complex multi-modal models to the extreme edge for time-constrained tasks such as autonomous navigation of robots and vehicles. 
    On small form-factor devices, e.g., nano-sized unmanned aerial vehicles (UAVs), such challenges are exacerbated by stringent constraints on energy efficiency and weight.
    In this paper, we explore embodied multi-modal AI-based perception for Nano-UAVs with the Kraken shield, a 7g multi-sensor (frame-based and event-based imagers) board based on Kraken, a \SI{22}{\nano\meter} SoC featuring multiple acceleration engines for multi-modal event and frame-based inference based on spiking (SNN) and ternary (TNN) neural networks, respectively. 
    Kraken can execute SNN real-time inference for depth estimation at 1.02k inf/s, \SI{18}{\micro\joule /inf}, TNN real-time inference for object classification at 10k inf/s,  \SI{6}{\micro\joule /inf}, and real-time inference for obstacle avoidance at \SI{221}{\fps}, \SI{750}{\micro\joule /inf}.
\end{abstract}

\vspace{-10pt}
\section{Introduction}
The future of artificial intelligence (AI) is embodied: use cases where inference on complex, multi-modal machine learning models cannot be offloaded to the cloud are bound to rapidly increase as value-added functions will be increasingly tied to autonomous, safe, and robust operation of devices, robots, and vehicles. Enabling multi-modal inference on autonomous devices requires addressing major efficiency improvements in circuits and systems, coupled with architectural evolution to ensure sustainable, secure, safe, and predictable operation. 

Aimed at addressing these staggering challenges, Kraken~\cite{KrakenHCS} is an ultra-low-power, heterogeneous \gls{soc} architecture integrating three acceleration engines and a vast set of IO peripherals to enable efficient interfacing with standard frame-based sensors and novel event-based \gls{dvss}~\cite{dvs}.
With the Kraken \gls{soc}, we take a step towards achieving autonomy on extremely constrained nano-sized unmanned aerial vehicles (UAVs). In this paper, we demonstrate multi-visual-sensor ML-based perception capabilities both on event-based and BW/RGB frame-based image sensors, and we show how to combine them to solve multi-modal perception tasks previously impossible on a single low-power chip~\cite{Skydio}.

We describe a custom-designed lightweight board hosting the SoC, sensors, optics, and power supply. This board, called Kraken shield, weights just 7g and can be easily carried by a nano-UAV, enabling acceleration of sparse event-driven \gls{snn}-based visual odometry tasks at \SI{1.02}{\kilo Inf/\second}, \SI{18}{\micro \joule /inf}. Moreover, it can perform TNN-based object classification at more than \SI{10}{\kilo Inf/\second}, \SI{6}{\micro \joule /inf}. Concurrently, it can run a DNN-based obstacle avoidance task at a rate of \SI{221}{\fps}, \SI{750}{\micro\joule/frame}, all within 5\% of the power envelope of a nano-UAV. 

\section{Architecture}

\begin{figure}[b]
    \vspace{-10pt}
    \centering
    \includegraphics[width=0.45\textwidth]{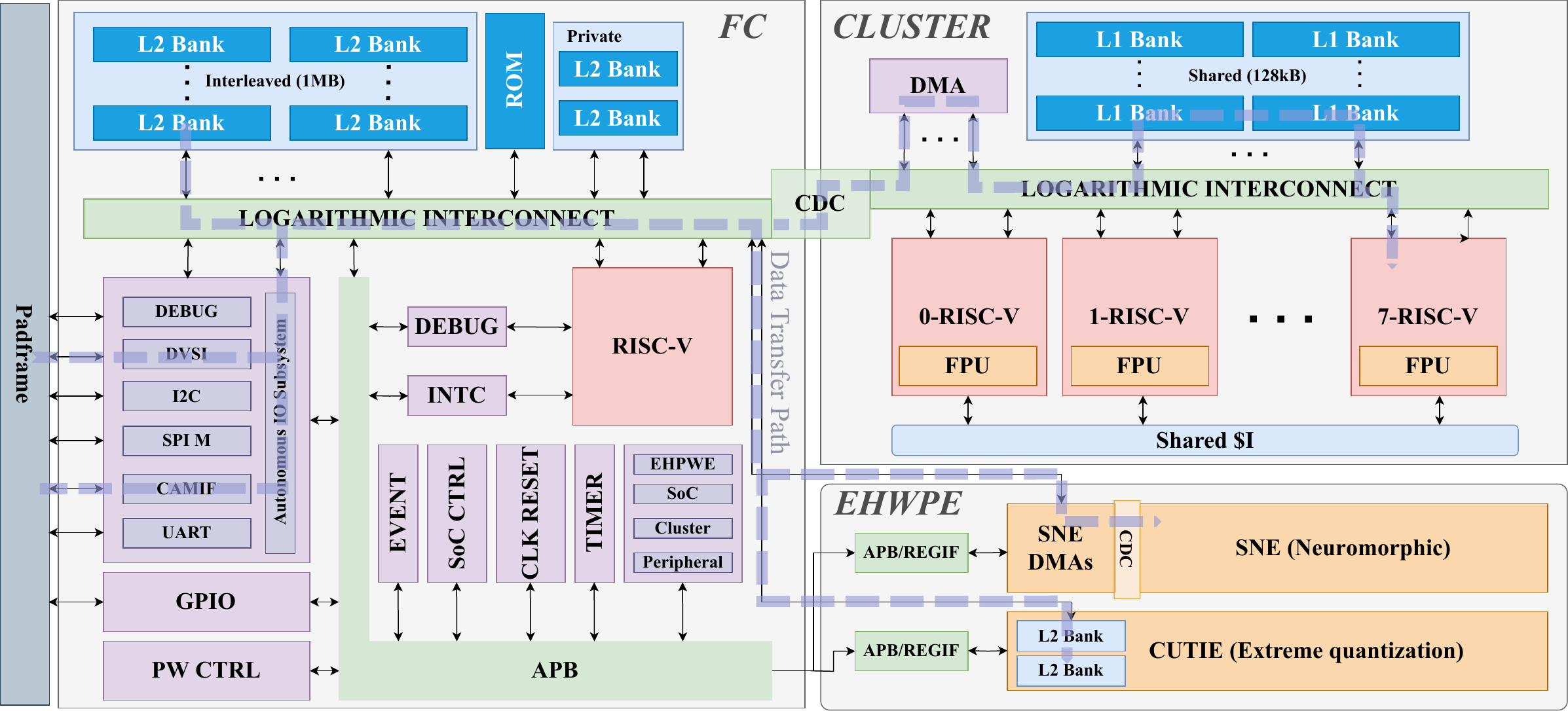}
    \caption{Architectural Block Diagram of the Kraken System on Chip (SoC). The diagram
    shows the FC, cluster, and accelerator domain hosting two accelerators, SNE and
    CUTIE for ML-based perception. The peripherals are centered around the FC, and
    data paths detailed in Section 4 are highlighted in a lilac shade. }
    \label{fig:syslevel}
    \vspace{-10pt}
\end{figure}

The Kraken SoC (Fig. \ref{fig:syslevel}) is built around a \SI{32}{\bit} fabric controller (FC) RISC-V core, it hosts \SI{1}{\mebi\byte} of scratchpad SRAM memory (L2) and a vast set of standard peripherals: 4 QSPI, 4 I2C, 2 UART and 48 GPIOs. Moreover, Kraken features 1 \gls{dvs} interface to acquire asynchronous stream of visual events, and 1 CPI interface to acquire frame-based images.
The FC can offload compute intensive kernels to three programmable, power-gateable accelerators.

\subsubsection{Sparse Neural Engine (SNE)}
\begin{figure}[tb]
\vspace{-10pt}
    \centering
    \begin{subfigure}[b]{0.9\linewidth}
        \centering
        \includegraphics[width=\linewidth]{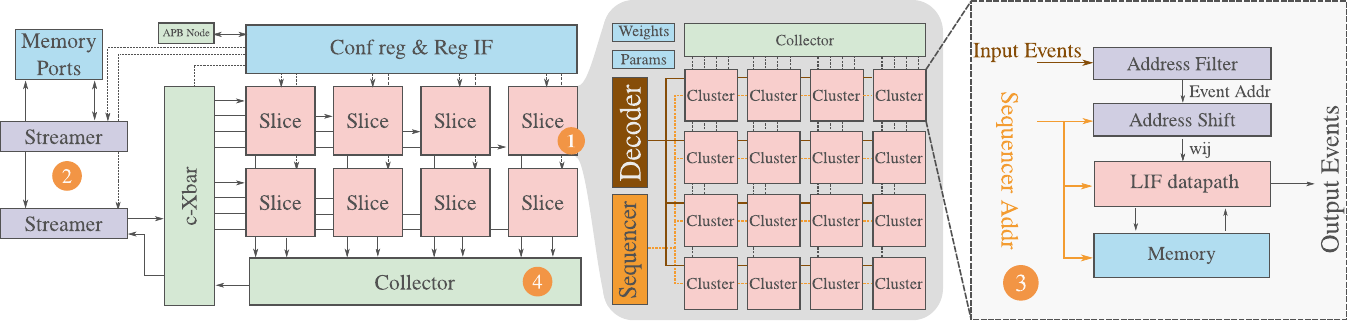}
        \caption{Data-path Schematic View of the Sparse Neural Engine (SNE). The diagram illustrates the time-domain multiplexed neural processing architecture, centralized event-driven crossbar system, streamers for efficient data handling, and clusters comprising LIF neurons.}
        \label{fig:sne_archi}
    \end{subfigure}
    
    \begin{subfigure}[b]{0.9\linewidth}
        \centering
        \includegraphics[width=\linewidth]{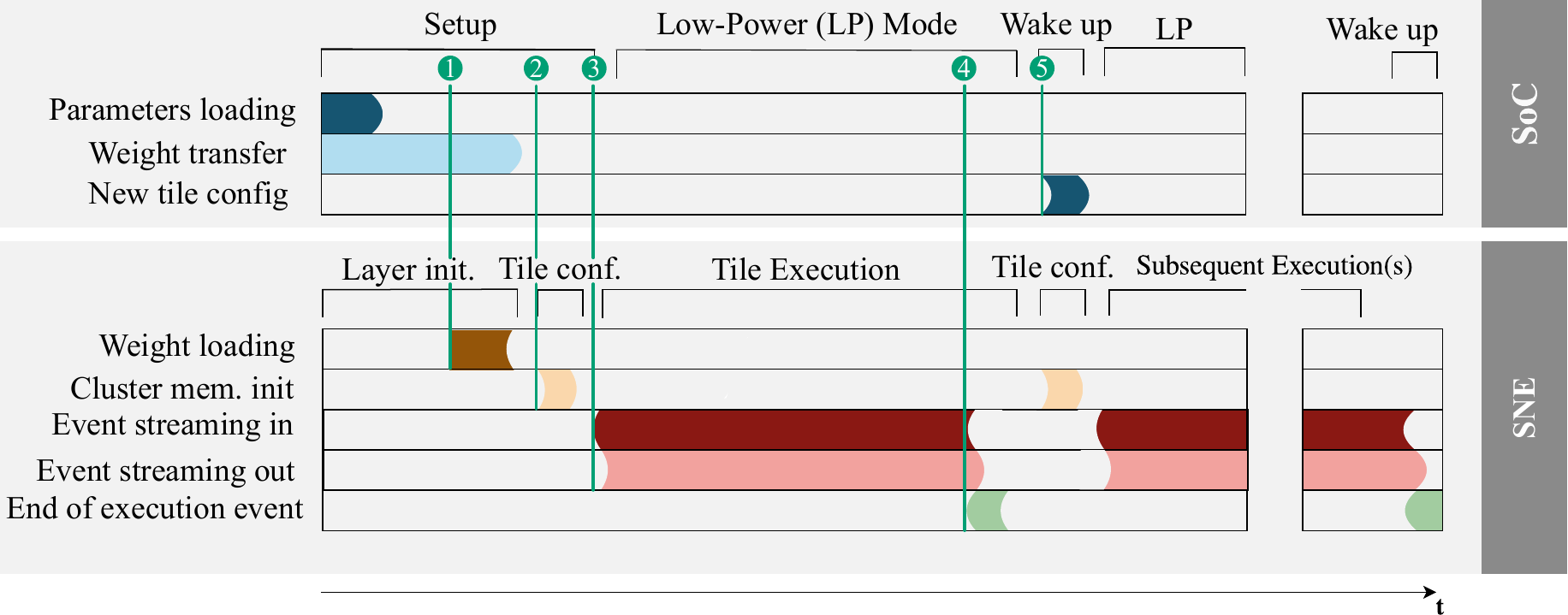}
        \caption{Scheduling diagram of the \texttt{SOC} (top) and \texttt{SNE} (bottom) for one layer inference. The neuron parameter and weight loading are executed once per layer. Once \texttt{SNE} starts its operation, the host is put into low-power mode until the tile is executed, and an end-of-execution event is issued to wake up the \texttt{SOC} \numberincircletiming{5}. Then, new tile parameters are loaded, and the streamers are re-programmed to load the events of a new tile.}
        \label{fig:sne_timing}
    \end{subfigure}
    
    \caption{Sparse Neural Engine (SNE) Architecture and Timing}
    \label{fig:sne_combined}
    \vspace{-10pt}
\end{figure}

The SNE is an inference engine designed to accelerate Spiking Convolutional Neural Networks (SCNNs). It supports 4-bit 3$\times$3 kernels for efficient computation and simulates neuronal activity with 8-bit leaky-integrate and fire (LIF) models. 
SNE's architecture is illustrated in Fig.~\ref{fig:sne_archi}. It features eight parallel processing engines called \emph{slices} \numberincircle{1}. Each connects to a central crossbar, linked to two programmable streamers \numberincircle{2}, operating as direct memory access (DMA) engines to load neural network weights \numberincircletiming{1}-\numberincircletiming{2} and stream events \numberincircletiming{3}-\numberincircletiming{4}. The streamers work with the re-programmable crossbar, implementing versatile data transfer schemes between the L2 memory and SNE slices. Supported dataflows include streaming events from L2 memory to multiple slices (input broadcast), streaming events between slices (internal redirection), or streaming events from multiple slices to the L2 memory (output streaming).

SNE accepts two kinds of events: \textit{i) Spike events}, in the "list of coordinates" (COO) format, containing \textit{x}, \textit{y}, and channel (\textit{c}) event coordinates, also employed by Kraken's event-based peripherals, such as DVSI; the DVS produces positive and negative events mapped to two channels of the COO format, \textit{ii) Time events}, i.e., 28-bit timestamps introduced between successive event-frames acquired from the DVS.

Slices contain 16 clusters \numberincircle{1}, each executing 64 concurrent LIF neuron functions using time-domain multiplexing (TDM) based single digital neuron data path. Every neuron state during TDM execution is stored in a local standard cell-based memory inside the cluster, directly connected to the datapath \numberincircle{3}.

SNE accelerates two types of SNN layers: 3$\times$3 convolutional layers (up to 256 input channels) and linear layers (maximum input dimension of 2304). Neurons are mapped on SNE's computational units in output stationary format. The weights and parameters of LIF neurons, including threshold potential or layer-wise exponential decay time constant, are pre-loaded via streamers into a shared buffer \numberincircle{1}, accessible by all clusters simultaneously.

Events are broadcast to all slices and internal clusters, where each cluster responds only to input events falling within the receptive field of the neurons mapped to that cluster. The LIF data path in each cluster can process two types of events: temporal events, transmitted to advance the SNN time and induce exponential decay in neuron potential, and spike events, leading to an increase or decrease of the neuron potential based on the respective input-to-output neuron weight.

The clusters use a highly efficient update mechanism where only 9 ($3\times3$ receptive field window) of 64 neurons respond to spike events by changing their potential. As the SNN timesteps progress, all 64 neurons must adjust their potential according to an exponential decay factor. SNE fuses neuron potential updates from spike and time events, performed in an event-driven fashion when a spike event is present. The \emph{time of last update} is registered per neuron and used to select a decay coefficient from a LUT, corresponding to the cumulative decay that would have happened from the \emph{time of last update} to the current time step. This saves 55/64 neuron update operations on each cluster at every time step, which balances silicon area and throughput. To implement the LIF function for a cluster of 64 neurons, SNE uses a single data path in TDM for 12 cycles: 1 cycle to update each of the nine activated neurons with the fused spike-time update and 3 cycles to restart the mechanism before processing the next input event.

The cluster operates in lockstep, orchestrated by a sequencer \numberincircle{3}. Multiple output events might be produced by the clusters in the same cycle. These are collected and transferred outside the slice through a buffering and arbitration unit called collector \numberincircle{3}. A similar mechanism merges the event streams of multiple slices \numberincircle{4}.

SNE is energy-proportional, i.e., the execution time of a layer-wise network inference is proportional to the number of input and output events of each layer, reflecting the network's sparsity. \figurename~\ref{fig:sne_timing} shows the timing diagram of the SNE tiled inference.

\subsubsection{Completely Unrolled Ternary Inference Engine (CUTIE)}
CUTIE is a Ternary Neural Network (TNN) accelerator designed to maximize energy efficiency during inference by minimizing data movement. All ternary weights are kept on-chip (in a compressed 1.6 bits/weight format), eliminating energy-intensive external data transfers. 

\begin{figure}[tb]
\vspace{-10pt}
    \centering
    \begin{subfigure}[b]{0.9\linewidth}
        \centering
    \includegraphics[width=\linewidth]{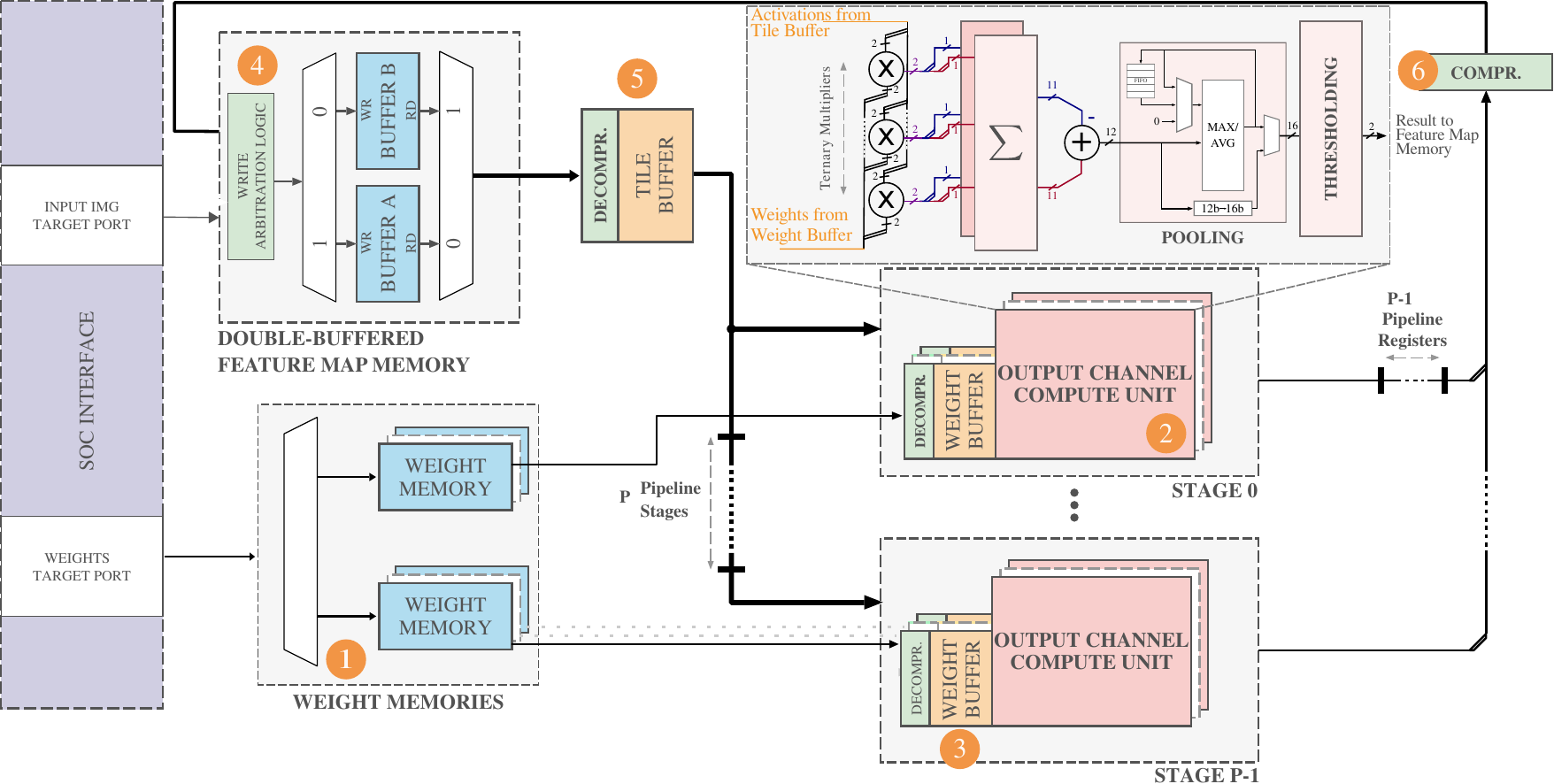}
    \caption{Data-path schematic view of CUTIE. The diagram shows the unrolled compute architecture, encoding/decoding blocks, storage for weights and feature maps, and the tile buffer module. The accelerator orchestrates data flow by buffering entire feature map windows in the tile buffer and then computing the convolution with pre-loaded weights in the OCUs. Eventually, the result is stored back in the feature map memory.}
    \label{fig:cutie_architecture}
    \end{subfigure}
    
    \begin{subfigure}[b]{0.9\linewidth}
        \centering
        \includegraphics[width=\linewidth]{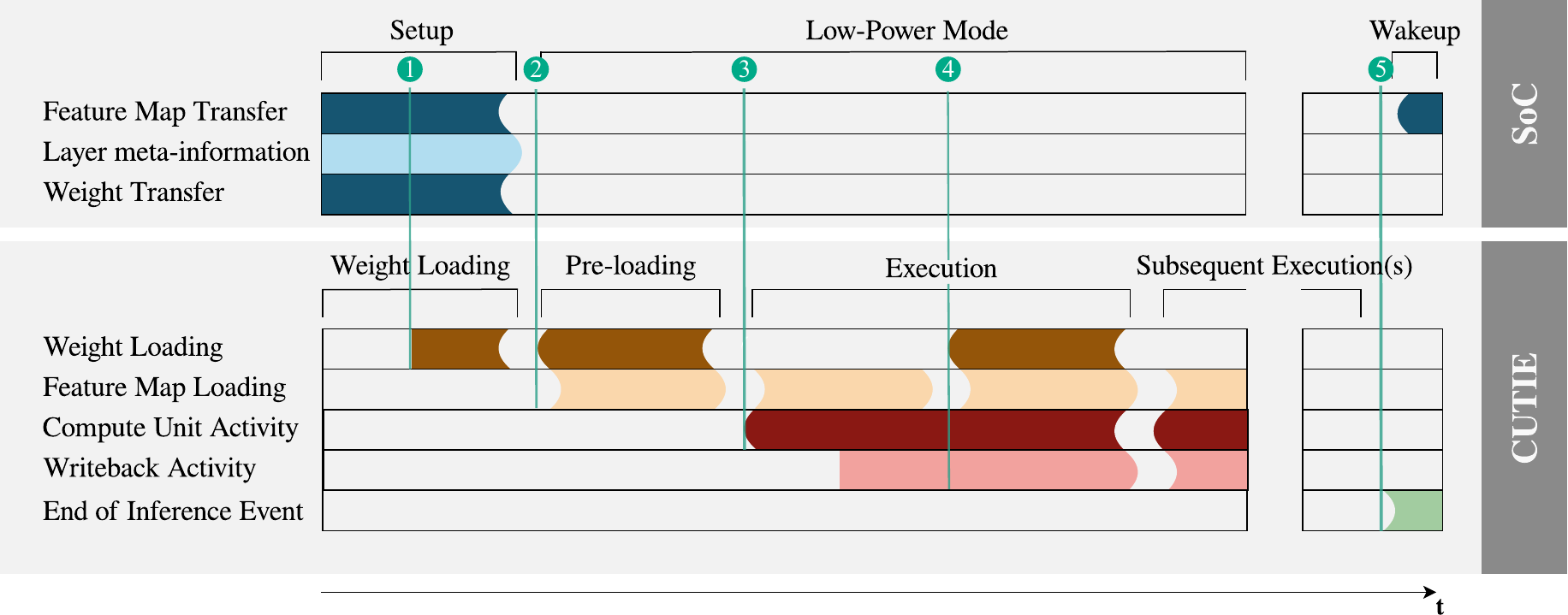}
        \caption{Scheduling diagram of the SOC (top) and CUTIE (bottom) for continuous inference. The setup and weight loading are only needed for the first layer. Subsequent loading phases are overlapped with the execution phase to hide the latency of scheduling a new layer to a single cycle. Once CUTIE starts its operation, the host is put into low-power mode since it carries all layer information inside its dedicated memory locations. }
        \label{fig:cutie_timing}
    \end{subfigure}
    
    \caption{Completely Unrolled Ternary Inference Engine (CUTIE) Architecture and Timing}
    \label{fig:cutie_combined}
    \vspace{-10pt}
\end{figure}

CUTIE's block and timing diagrams are shown in \figurename~\ref{fig:cutie_architecture} and \figurename~\ref{fig:cutie_timing}. CUTIE leverages TNN memory efficiency, where weights and activations compress to less than 2 bits per symbol \numberincircle{6}. Including a zero state reduces switching activity by \emph{turning off} network weight connections, thus lowering energy use at equivalent memory usage~\cite{tnn_zero_skipping}. Compressed neural network parameters are decompressed \numberincircle{5} while streamed into the OCUs \numberincircle{3}, \numberincircletiming{2}, using an efficient scheme grouping values into sets of five, stored in 8 bits to achieve a 1.6 bits per symbol compression rate~\cite{tnn_compression}.

CUTIE supports ternary weight convolutional layers with optional pooling, using two thresholds for ternarizing outcomes. Each layer's weights are consecutively stored in weight memories \numberincircle{1}. The OCUs \numberincircle{2} with dual latch-based buffers for filter weights facilitate output pixel computation and layer overlap \numberincircletiming{1}. Simultaneous weight loading for upcoming layers and current layer computation \numberincircletiming{2} - \numberincircletiming{4} optimize efficiency. The \emph{end of inference} signal indicates to the host controller that the final class scores are available and valid, and CUTIE is ready to receive the next feature map input \numberincircletiming{5}. The OCU's ternary MAC unit calculates single weight and activation products, summing them into a 16-bit value. This value is passed through a threshold decider, producing 2-bit outputs, accumulated to five output values, and compressed into an 8-bit binary representing five trits, optimizing data representation.

Neural networks are executed layer-wise on CUTIE, featuring a flat design hierarchy. This allows channel-wise output feature maps computation, with 96 OCUs, amounting to 82944 ternary MAC/cycle ($3\times3$ convolutions). A private memory for each OCU \numberincircle{3} buffers weights to reduce energy consumption spent on data transfers, while shared feature map storage \numberincircle{4} across OCUs optimizes activation data reuse. A tile buffer \numberincircle{5} creates square windows of the input feature map using sliding windows, which are then broadcast to the OCUs. This allows the processing of 24 million frames per second for the CIFAR-10 dataset ($32\times 32$ pixels), achieving throughput up to 5.4 TOp/s. 

\subsubsection{Parallel ultra-low power cluster (PULP)}
The cluster (CL) comprises eight 32-bit RISC-V cores, supporting the RV32-IMCF and XpulpNN~\cite{ref:garofalo} ISA extensions. These include hardware support for fast loop programming, standard and narrow-bit MAC operations, and SIMD instructions tailored for low-precision ML workloads down to 2-bits. Each core has a private FPU, useful for near-sensor tasks that cannot be easily quantized, like sensor data conditioning, filtering, and control tasks.
The cluster features an L1 TCDM with 128 \si{\kilo\byte} capacity, a banking factor of two, and word-level address interleaving~\cite{capotondi_2016}. Up to eight concurrent memory requests with single-cycle access latency and a banking conflict probability generally lower than 10\% can be handled. 

The cluster incorporates a dedicated \emph{Hardware Synchronizer} supporting fast event management, parallel thread dispatching, and synchronization~\cite{glaser_2021}. This enables fine-grained parallelism and high energy efficiency in parallel workloads. The cluster can be clock-gated with a single core granularity, reducing the dynamic power consumption when synchronizing cores.

\begin{figure*}[t]
    \vspace{-5pt}
    \centering
    \begin{subfigure}[b]{0.54\linewidth}
        \includegraphics[width=\linewidth]{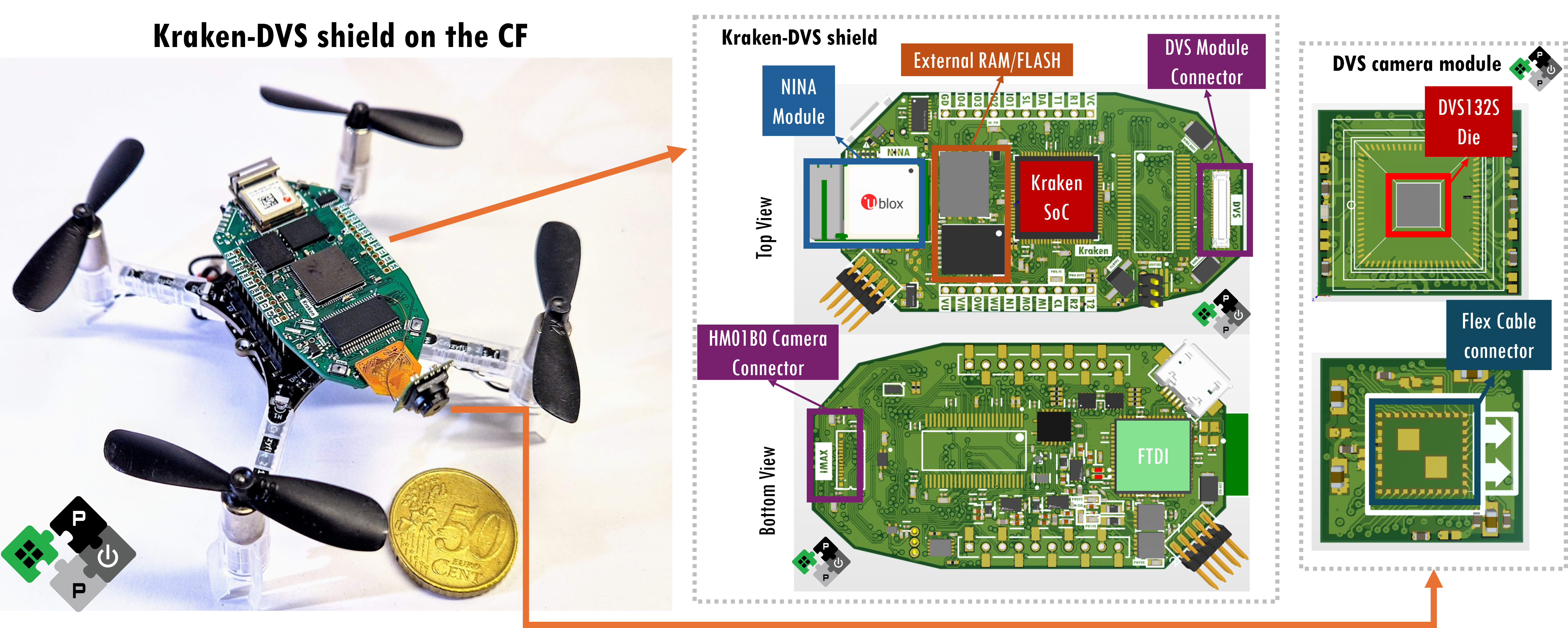}
        \caption{Shield mounted on a CF2.1 with Kraken-DVS Shield and DVS camera module 3D renderings}
        \label{fig:shield-module}
    \end{subfigure}
    \begin{subfigure}[b]{0.17\linewidth}
        \includegraphics[width=\linewidth]{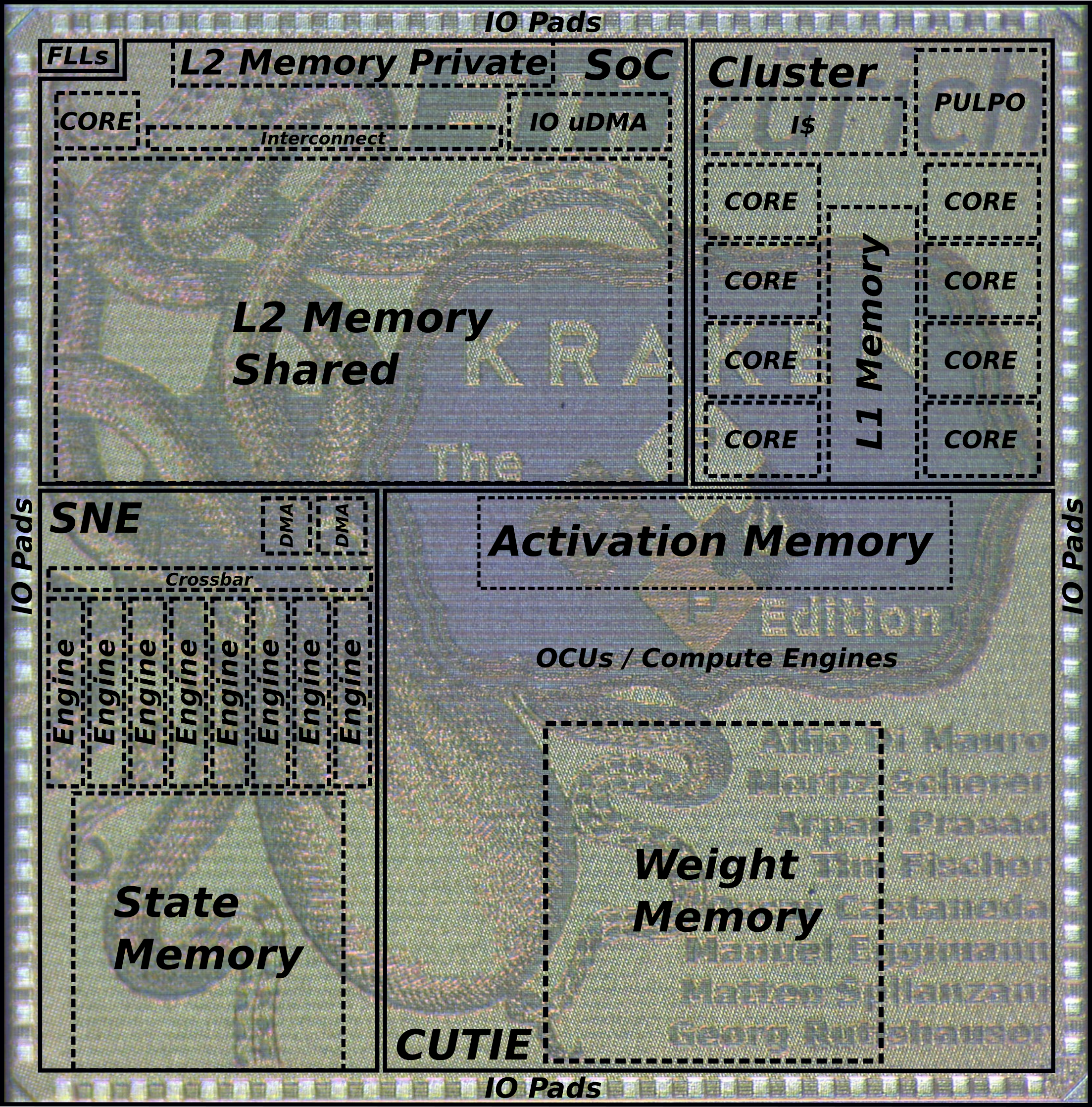}
        \caption{Die micrograph of the Kraken SoC showing the different power domains.}
        \label{fig:dieshot}
    \end{subfigure}
    
    \caption{Complete system overview}
    \label{fig:cutie_combined}
    \vspace{-10pt}
\end{figure*}

\section{AI Nano-UAV Shield}

While the accelerated computing capabilities of advanced SoCs like Kraken certainly provide considerable performance enhancements, their ultimate deployment, and usability on platforms constrained by size and weight, such as nano-UAVs, is only feasible if such an SoC can effectively interact with external off-the-shelf modules and components, to eventually build a complete low-power AI-capable on-board computer (OBC). 
Based on the Crazyflie (CF) 2.1\footnote{\url{https://www.bitcraze.io/products/crazyflie-2-1/}}, a modular open-source 27g, 10cm nano-UAV platform, we have developed a compact and lightweight pluggable module for capturing and processing event-based images. The platform, depicted in \figurename \ref{fig:shield-module}, encompasses:

\subsubsection{CF-pluggable 50x27mm PCB Shield}  housing the Kraken SoC, an external 512Mb Flash memory and a 512Mb/64Mb HyperFlash/HyperRAM, extending Kraken storage, a WiFi module (NINA) for wireless data transmission (40Mbps), an FTDI chip, voltage regulators, and connectors for both the DVS and an 8-bit parallel QVGA grayscale camera (HM01B0).

\subsubsection{DVS Camera Module} A Chip-on-board (COB) module integrating the DVS132S die bonded to an interface PCB, including decoupling capacitors and a flex cable interface to connect the main board. 
The module also integrates a miniaturized lens with an M4 mount type directly glued to the PCB.
The lens has a field-of-view (FoV) of \SI{66}{\degree}$\times$\SI{51}{\degree}, and a diameter of \SI{5}{\milli\meter}, while the side of the squared lens holder is \SI{6.3}{\milli\meter}.
\textcolor{black}{The complete perception system, including the shield with Kraken, DVS, and grayscale cameras, has a total weight of \SI{7}{\gram}}, compatible with the payload of a nano-UAV.

\section{Results}
In this section, we present measurements to assess the energy consumption of Kraken's subsystems, presented in the previous sections. Kraken has been fabricated in \SI{22}{\nano\meter} FDX technology. The maximum frequency achieved on the SoC is 300MHz at 0.8V. Fig. \ref{fig:dieshot} shows a micrograph of the chip.
Each subsystem has been benchmarked under end-to-end tasks that are essential functional blocks for autonomous navigation on nano-UAVs. Specifically, SNN-based depth estimation, TNN-based object classification, and DNN-based obstacle avoidance are executed on SNE, CUTIE, and the PULP cluster, respectively.

SNE's energy consumption has been measured on a 4bit quantized, 6-layer, low-memory footprint, Convolutional Spiking Neural Network (CSNN) with a topology analogous to the LIF-FireNet proposed in \cite{NEURIPS2021_39d4b545}, a neural network used to solve a task of optical flow reconstruction from DVS events. During inference, the \gls{sne} consumes 98mW at 220MHz, 0.8V. At this frequency, \gls{sne} can perform more than \SI{20.8}{\kilo Inf/\second} at low (1\%) network activity, and \SI{1.02}{\kilo Inf/\second} at the 20\% average activity. The energy per inference is comprised between \SI{7.5}{\micro \joule /inf} and \SI{18}{\micro \joule /inf} at 1\% and 20\% network activity, respectively.

CUTIE's Energy consumption has been measured while executing a ternary CIFAR10 network derived from \cite{Moons2018} on an object classification task. \textcolor{black}{During inference, CUTIE consumes 110mW, at 0.8V, when clocked at 330MHz.} At this frequency CUTIE can execute more than \SI{10}{\kilo Inf/\second}. The energy per inference is \SI{6}{\micro \joule /inf}.

\begin{figure}[tb]
    \centering
    \includegraphics[width=0.9\linewidth]{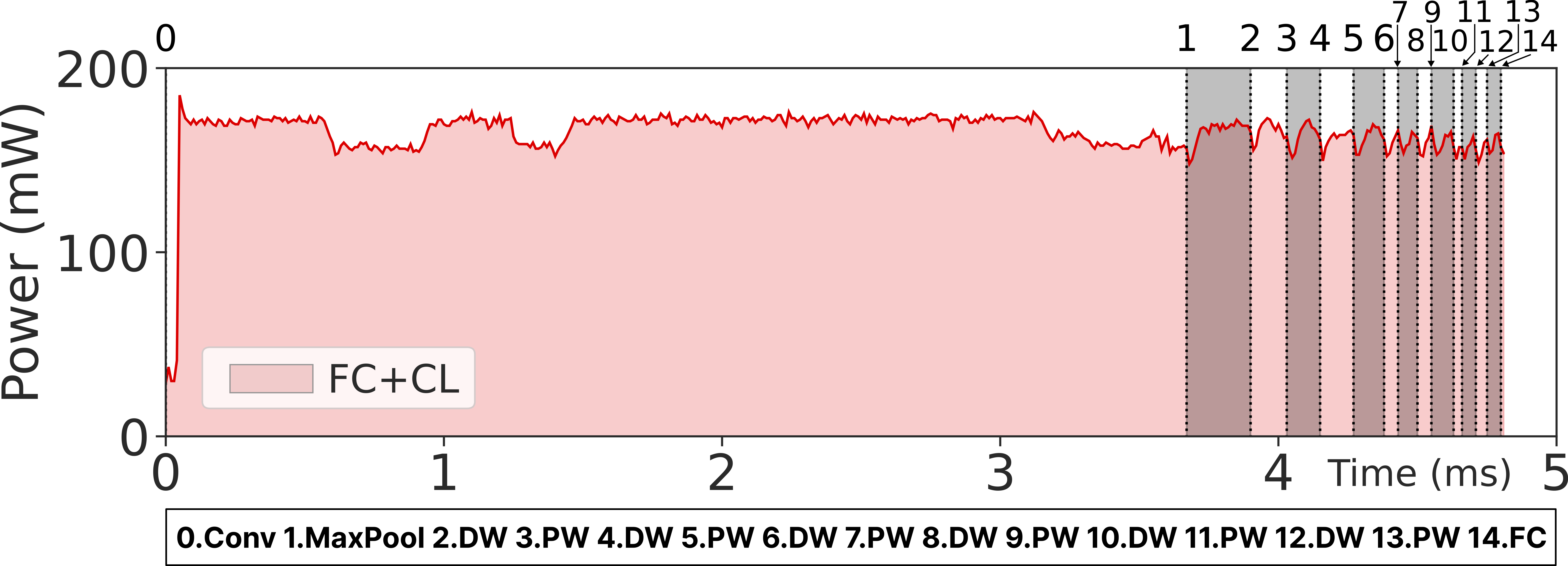}
    \caption{Kraken power waveform executing Tiny-PULP-Dronet at FC@\SI{280}{\mega\hertz}, CL@\SI{300}{\mega\hertz}, Vdd@\SI{0.8}{\volt}.
    }
    \label{fig:dronet}
    \vspace{-10pt}
\end{figure}

The PULP cluster performs a navigation and obstacle avoidance task based on an 8-bit quantized Tiny-PULP-Dronet network presented in\cite{tiny_dronet}. 
Using the maximum performance configuration, i.e., FC@\SI{280}{\mega\hertz}, CL@\SI{300}{\mega\hertz}, Vdd@\SI{0.8}{\volt}, the network executes at a rate of \SI{211}{\fps} in an \SI{165}{\milli\watt} power envelope, achieving an energy per inference of \SI{750}{\micro\joule/frame}. The power envelope of a single inference is reported in \figurename ~\ref{fig:dronet}, highlighting the execution of each layer of the CNN, where DW and PW are depthwise and pointwise convolutions, respectively.
When setting the SoC at the energy-efficient configuration, i.e., FC@\SI{110}{\mega\hertz}, CL@\SI{110}{\mega\hertz}, Vdd@\SI{0.55}{\volt}, the network executes at a rate of \SI{71}{\fps} in an \SI{23}{\milli\watt} power envelope, achieving an energy per inference of \SI{322}{\micro\joule/frame}, 2.3$\times$ more efficient than the maximum performance configuration.

\gls{sne} positions itself in a different Pareto point compared to the work of Tan et al.~\cite{tan40nm89pJSOP2023}, reporting 1.4$\times$ higher Energy/SOp but also \textcolor{black}{7.5$\times$} higher throughput. 
With respect to Wang et al.~\cite{wang7pJSOPNeuromorphic2023}, \gls{sne} achieves 1.3$\times$ lower Energy/SOp.

\acrshort{cutie} achieves the lowest energy per inference on the CIFAR-10 dataset, outperforming Knag et al. ~\cite{knag617TOPSAllDigitalBinary2021} by 1.1 $\times$ and Ryu et al. \cite{ryuBinarywareHighPerformanceDigital2023} by 1.18 $\times$. Compared to Cheon et al. \cite{cheon2941TOPSChargeDomain10T2023}, \acrshort{cutie} positions itself in a different Pareto point, reporting 2.8$\times$ lower energy efficiency but 6.4$\times$ higher throughput.

Kraken programmable RISC-V cluster achieves comparable efficiency with respect to SoA PULP systems~\cite{rossiVegaTenCoreSoC2022} while boosting performance by a factor of 2.7$\times$ on 8-bit workloads. Moreover, Kraken improves the PULP architectures reported in ~\cite{rossiVegaTenCoreSoC2022} by supporting low-precision SIMD operations. On 2-bit workloads, it achieves a peak computational efficiency of 1.6 TOPS/W, which, to the best of our knowledge, is the highest efficiency reported by RISC-V cores on quantized DNN inference~\cite{dustin}.

\section{Outlook and conclusion}
Kraken and its shield are a first demonstration of a perception system capable of executing multi-modal Embodied ML models for autonomous nano-UAVs. It runs real-time SNN inference for depth estimation at 1.02k inf/s at just 18 uJ/inf. At the same time, It can execute real-time TNN inference at 10k inf/s, consuming only 6 uJ/inf. Moreover, it can perform obstacle avoidance in real-time, at \SI{221}{\fps}, \SI{750}{\micro\joule /inf}. All multi-modal perception tasks can run concurrently within a power envelope of \SI{373}{\milli \watt}, 5.3\% of a nano-UAV's total power budget. 
Kraken is a first step toward embodied AI for tightly power and size-constrained autonomous systems. Much work lies ahead to move from perceptive AI to generative AI under a similar power envelope. Achieving this objective will enable much higher levels of intelligence and autonomy, even on tiny autonomous systems, ultimately approaching what nature has achieved with insects in millions of years of evolution.

\bibliography{IEEEabrv,./bibliography}

\begin{thebibliography}{10}
\providecommand{\url}[1]{#1}
\csname url@samestyle\endcsname
\providecommand{\newblock}{\relax}
\providecommand{\bibinfo}[2]{#2}
\providecommand{\BIBentrySTDinterwordspacing}{\spaceskip=0pt\relax}
\providecommand{\BIBentryALTinterwordstretchfactor}{4}
\providecommand{\BIBentryALTinterwordspacing}{\spaceskip=\fontdimen2\font plus
\BIBentryALTinterwordstretchfactor\fontdimen3\font minus \fontdimen4\font\relax}
\providecommand{\BIBforeignlanguage}[2]{{%
\expandafter\ifx\csname l@#1\endcsname\relax
\typeout{** WARNING: IEEEtran.bst: No hyphenation pattern has been}%
\typeout{** loaded for the language `#1'. Using the pattern for}%
\typeout{** the default language instead.}%
\else
\language=\csname l@#1\endcsname
\fi
#2}}
\providecommand{\BIBdecl}{\relax}
\BIBdecl

\bibitem{KrakenHCS}
A.~Di~Mauro \emph{et~al.}, ``Kraken: A direct event/frame-based multi-sensor fusion soc for ultra-efficient visual processing in nano-uavs,'' in \emph{2022 IEEE Hot Chips 34 Symposium (HCS)}, 2022, pp. 1--19.

\bibitem{dvs}
C.~Li \emph{et~al.}, ``A 132 by 104 $10\mu m$-pixel $250\mu w$ 1kefps dynamic vision sensor with pixel-parallel noise and spatial redundancy suppression,'' in \emph{2019 Symposium on VLSI Circuits}, 2019, pp. C216--C217.

\bibitem{Skydio}
A.~Bachrach, ``Skydio autonomy engine: Enabling the next generation of autonomous flight,'' in \emph{2021 IEEE Hot Chips 33 Symposium (HCS)}, 2021.

\bibitem{tnn_zero_skipping}
T.~Na, ``Ternary output binary neural network with zero-skipping for mram-based digital in-memory computing,'' \emph{{IEEE} Trans. Circuits Syst. {II}}, vol.~70, no.~7, pp. 2655--2659, 2023.

\bibitem{tnn_compression}
O.~Muller \emph{et~al.}, ``Efficient decompression of binary encoded balanced ternary sequences,'' \emph{{IEEE} Trans. {VLSI} Syst.}, vol.~27, no.~8, pp. 1962--1966, 2019.

\bibitem{ref:garofalo}
A.~Garofalo \emph{et~al.}, ``Xpulpnn: Enabling energy efficient and flexible inference of quantized neural networks on risc-v based iot end nodes,'' \emph{{IEEE} Trans. Emerg. Topics Comput.}, vol.~9, no.~3, pp. 1489--1505, 2021.

\bibitem{capotondi_2016}
A.~Capotondi \emph{et~al.}, ``On the effectiveness of openmp teams for cluster-based many-core accelerators,'' \emph{HPCS}, pp. 667--674, 2016.

\bibitem{glaser_2021}
F.~Glaser \emph{et~al.}, ``Energy-efficient hardware-accelerated synchronization for shared-l1-memory multiprocessor clusters,'' \emph{{IEEE} Trans. Parallel Distrib. Syst.}, vol.~32, no.~3, pp. 633--648, 2021.

\bibitem{NEURIPS2021_39d4b545}
J.~Hagenaars \emph{et~al.}, ``Self-supervised learning of event-based optical flow with spiking neural networks,'' in \emph{Advances in Neural Information Processing Systems}, M.~Ranzato \emph{et~al.}, Eds., vol.~34.\hskip 1em plus 0.5em minus 0.4em\relax Curran Associates, Inc., 2021, pp. 7167--7179.

\bibitem{Moons2018}
B.~{Moons} \emph{et~al.}, ``Binareye: An always-on energy-accuracy-scalable binary cnn processor with all memory on chip in 28nm cmos,'' in \emph{Proc. IEEE CICC}, 2018.

\bibitem{tiny_dronet}
L.~Lamberti \emph{et~al.}, ``Tiny-pulp-dronets: Squeezing neural networks for faster and lighter inference on multi-tasking autonomous nano-drones,'' in \emph{2022 IEEE 4th International Conference on Artificial Intelligence Circuits and Systems (AICAS)}, 2022, pp. 287--290.

\bibitem{tan40nm89pJSOP2023}
P.-Y. Tan \emph{et~al.}, ``A 40-nm 1.89-{{pJ}}/{{SOP Scalable Convolutional Spiking Neural Network Learning Core With On-Chip Spatiotemporal Back-Propagation}},'' \emph{{IEEE} Trans. {VLSI} Syst.}, vol.~31, no.~12, pp. 1994--2007, Dec. 2023.

\bibitem{wang7pJSOPNeuromorphic2023}
B.~Wang \emph{et~al.}, ``1.{{7pJ}}/{{SOP Neuromorphic Processor}} with {{Integrated Partial Sum Routers}} for {{In-Network Computing}},'' in \emph{2023 {{IEEE International Symposium}} on {{Circuits}} and {{Systems}} ({{ISCAS}})}.\hskip 1em plus 0.5em minus 0.4em\relax Monterey, CA, USA: IEEE, May 2023, pp. 1--5.

\bibitem{knag617TOPSAllDigitalBinary2021}
P.~C. Knag \emph{et~al.}, ``A 617-{{TOPS}}/{{W All-Digital Binary Neural Network Accelerator}} in 10-nm {{FinFET CMOS}},'' \emph{{IEEE} J. Solid-State Circuits}, vol.~56, no.~4, pp. 1082--1092, Apr. 2021.

\bibitem{ryuBinarywareHighPerformanceDigital2023}
S.~Ryu \emph{et~al.}, ``Binaryware: {{A High-Performance Digital Hardware Accelerator}} for {{Binary Neural Networks}},'' \emph{{IEEE} Trans. {VLSI} Syst.}, vol.~31, no.~12, pp. 2137--2141, Dec. 2023.

\bibitem{cheon2941TOPSChargeDomain10T2023}
S.~Cheon \emph{et~al.}, ``A 2941-{{TOPS}}/{{W Charge-Domain 10T SRAM Compute-in-Memory}} for {{Ternary Neural Network}},'' \emph{{IEEE} Trans. Circuits Syst. {I}}, vol.~70, no.~5, pp. 2085--2097, May 2023.

\bibitem{rossiVegaTenCoreSoC2022}
D.~Rossi \emph{et~al.}, ``Vega: {{A Ten-Core SoC}} for {{IoT Endnodes With DNN Acceleration}} and {{Cognitive Wake-Up From MRAM-Based State-Retentive Sleep Mode}},'' \emph{{IEEE} J. Solid-State Circuits}, vol.~57, no.~1, pp. 127--139, Jan. 2022.

\bibitem{dustin}
G.~Ottavi \emph{et~al.}, ``Dustin: A 16-cores parallel ultra-low-power cluster with 2b-to-32b fully flexible bit-precision and vector lockstep execution mode,'' \emph{{IEEE} Trans. Circuits Syst. {II}}, vol.~70, no.~6, pp. 2450--2463, 2023.

\end{thebibliography}

\end{document}